\documentclass[9pt,twocolumn]{extarticle}
\pdfoutput=1
\usepackage{soul}
\usepackage{titlesec} 

\usepackage[superscript, nomove]{cite}

\usepackage{float}
\usepackage{caption}
\usepackage{lipsum,ulem}
\usepackage[verbose]{placeins}

\usepackage{dsfont}
\usepackage{graphicx}
\usepackage{subcaption}
\usepackage[margin=0.9in]{geometry}
\usepackage[usenames,dvipsnames]{xcolor}
\usepackage[colorlinks,linkcolor=Blue,urlcolor=Blue,citecolor=Blue]{hyperref}
\usepackage{amsmath,amssymb}
\usepackage[squaren]{SIunits}

\usepackage{mathpazo}
\usepackage{courier}
\normalfont
\usepackage[T1]{fontenc}
\usepackage{caption}
\renewcommand{\thefigure}{\arabic{figure}}
\renewcommand{\figurename}{Fig.}
\usepackage{multirow}

\usepackage{upgreek}

\newcommand{\ad}[1]{\textsuperscript{#1}\kern-2pt}

\makeatletter
\def\blx@maxline{77}
\makeatother

\usepackage{capt-of}

\def\mytitle{A two-dimensional semiconductor-semimetal drag hybrid}

\title{\vspace{-1.0cm}\Huge\textbf{\textrm{\mytitle}}}  

\author{Yingjia Liu,$^{1,2,3}$ Kaining Yang,$^{4,5}$ Kenji Watanabe,$^{6}$ Takashi Taniguchi,$^{7}$ Wencai Ren,$^{1, 2\dagger}$ Zheng Vitto Han,$^{3,4,5\dagger}$ and Siwen Zhao$^{3\dagger}$}

\date{} 
\begin{document}
	\twocolumn[
		\maketitle 
		\vspace{-5mm}
		\begin{center}
			\begin{minipage}{1\textwidth}
				\begin{center}
					\textit{
					\\\textsuperscript{1} Shenyang National Laboratory for Materials Science, Institute of Metal Research, Chinese Academy of Sciences, Shenyang 110016, China
					\\\textsuperscript{2} School of Material Science and Engineering, University of Science and Technology of China, Anhui 230026, China
					\\\textsuperscript{3} Liaoning Academy of Materials, Shenyang 110167, P. R. China
					\\\textsuperscript{4} State Key Laboratory of Quantum Optics Technologies and Devices, Institute of Optoelectronics, Shanxi University, Taiyuan 030006, P. R. China
					\\\textsuperscript{5} Collaborative Innovation Center of Extreme Optics, Shanxi University, Taiyuan 030006, P. R. China
					\\\textsuperscript{6} Research Center for Electronic and Optical Materials, National Institute for Materials Science, 1-1 Namiki, Tsukuba 305-0044, Japan
					\\\textsuperscript{7} Research Center for Materials Nanoarchitectonics, National Institute for Materials Science,  1-1 Namiki, Tsukuba 305-0044, Japan
					\vspace{1.0cm}
					\\{$\dagger$} Corresponding to: wcren@imr.ac.cn, vitto.han@gmail.com, siwenzhao0126@gmail.com}
				\end{center}
			\end{minipage}
		\end{center}
\vspace{1.0cm}
\begin{abstract} 
\textbf{Lateral charge transport of a two-dimensional (2D) electronic system can be much influenced by feeding a current into another closely spaced 2D conductor, known as the Coulomb drag phenomenon -- a powerful probe of electron-electron interactions and collective excitations. Yet the materials compatible for such investigations remain limited to date. Especially, gapped 2D semiconductors with inherently large correlations over a broad gate range have been rarely accessible at low temperatures. Here, we show the emergence of a large drag response (drag resistance $R_\mathrm{drag}$ at the order of k$\Omega$, with a passive-to-active drag ratio up to $\sim$ 0.6) in a semiconductor-semimetal hybrid, realized in a graphene-MoS$_{2}$ heterostructure isolated by an ultrathin 3 nm hexagonal boron nitride (h-BN) dielectric. We observe a crossover of $T$- to $T^{2}$-dependence of $R_\mathrm{drag}$, separated by a characteristic temperature $T_{d} \sim E_{F}/k_{F}d$ ($d$ being the interlayer distance), in echo with the presence of a metal-insulator transition in the semiconducting MoS$_{2}$. Interestingly, the current nanostructure allows the decoupling of intralayer interaction-driven drag response by varying density in one layer with that in the other layer kept constant. A large Wigner–Seitz radius $r_{s}$ ($>$ 10 within the density range of 1 to 4$\times 10^{12} \mathrm{cm}^{-2}$) in the massive Schrödinger carriers in MoS$_{2}$ is thus identified to dominate the quadratic dependence of total carriers in the drag system, while the massless Dirac carriers in graphene induce negligible drag responses as a function of carrier density. Our findings establish semiconductor-semimetal hybrid as a platform for studying unique interaction physics in Coulomb drag systems.}
\end{abstract} 
] 

\clearpage
\newpage
	
\section*{Introduction}

In two closely spaced low dimensional conductors, charge carriers driving in one active layer is often observed to induce drag characteristics in another passive layer, yielding a current or voltage in the latter. Such effects offer a fundamental yet direct probe for electronic momentum and/or energy exchange via long range Coulomb interactions, as well as many-body physics beyond single-particle transport\cite{narozhny2016coulomb}. Indeed, Coulomb drag phenomena have been extensively manifested in different regimes, including quantum wells or graphene separated with large distance in the weak coupling limit \cite{jauho1993coulomb,gorbachev2012strong}, and excitonic condensation when interlayer charge carriers are matched in the quantum Hall limit \cite{kellogg2003bilayer, eisenstein2004bose, kellogg2002observation, kellogg2004vanishing, tutuc2004counterflow, nandi2012exciton, liu2017quantum,liu2022crossover,li2017excitonic}. More recently, emerging physical phenomena are also reported in exotic drag between graphene and superconductors, topological insulators, 1D-1D Luttinger liquid, quantum dots and mixed dimensional electrons\cite{tao2023josephson,du2021coulomb,anderson2021coulomb,laroche20141d, tabatabaei2020andreev, mitra2020anomalous}.

Among those reported, gapped two-dimensional (2D) semiconductors, with inherently large correlations in the massive carriers, have been a missing piece in the jigsaw puzzle of various drag regimes. Especially, a peculiar family of massive-massless double layers has remained largely unvisited. Taking the Wigner–Seitz radius $r_{s}$ (strong correlation when $r_{s}$ > 10) as a measure of interaction strength in 2D electron systems, massless Dirac fermions in monolayer graphene has a density-independent value of $r_{s} \sim$ 0.7 - 0.8\cite{das2011electronic,das2007many}. Meanwhile, in bilayer graphene and conventional 2D electron gases in quantum wells, Fermi surfaces are well defined and $r_{s}$ is sufficiently large only when carrier density is remained ultra low (< $10^{10}$ $\textrm{cm}^{-2}$), which is manifested in such as an unconventional negative frictional drag in the vicinity of charge neutral in double graphene bilayers\cite{lee2016giant,li2016negative}. Gapped 2D semiconductors, the transition metal dichalcogenides (TMDs) for instance, host strong tunable Coulomb interactions with $r_{s} >$ 10 across a broad gate range\cite{lin2019determining,ahn2023density}. The interplay of these massive interacting Schrödinger fermions with massless Dirac fermions in the Coulomb drag paradigm is expected to unveil new transport regimes, yet its experimental access has been rare, so far\cite{scharf2012coulomb,principi2012plasmons,gamucci2014anomalous}. This is mainly due to the grand challenge of obtaining Ohmic contacts and maintaining high-mobility charge transport at their low temperature ground states. 

In this work, we demonstrate large Coulomb drag responses in a semiconductor-semimetal hybrid, realized in a MoS$_{2}$–graphene heterostructure separated by an ultrathin 3 nm h-BN dielectric. Using a 2D window contact method, Ohmic contacts are realized in MoS$_{2}$ throughout the temperature range tested in this study. Unlike conventional drag systems, we observe a $R_{\textrm{drag}}$ as high as several hundred $\Omega$, with a passive-to-active drag ratio (PADR) reaching $\sim$ 0.6, orders of magnitude larger than previously reported values\cite{duan1993supercurrent,huang1995observation}. Furthermore, we identify a crossover in temperature dependence of $R_{\textrm{drag}}$, transitioning from a linear $T$-dependence at high temperatures to a quadratic $T$-dependence below a characteristic temperature $T_{d}$, which coincides with the onset of a metal-insulator transition in MoS$_{2}$. Our study further reveals the ability to decouple intralayer and interlayer correlation effects by independently tuning the carrier density in one layer while keeping the other fixed. It is found that the interacting massive Schrödinger fermions in MoS$_{2}$ dictates the quadratic dependence of total carriers in the drag signal, while the massless Dirac electrons in graphene contribute negligible drag responses. These observations highlight the crucial role of intralayer correlations in MoS$_{2}$ in amplifying the drag responses. Our findings not only expand the scope of Coulomb drag studies to correlated 2D semiconductors but also offer insights into designing next-generation interaction-driven electronic devices.

\section*{Results}

 \begin{figure*}[ht!]
	\centering
	\includegraphics[width=0.9\linewidth]{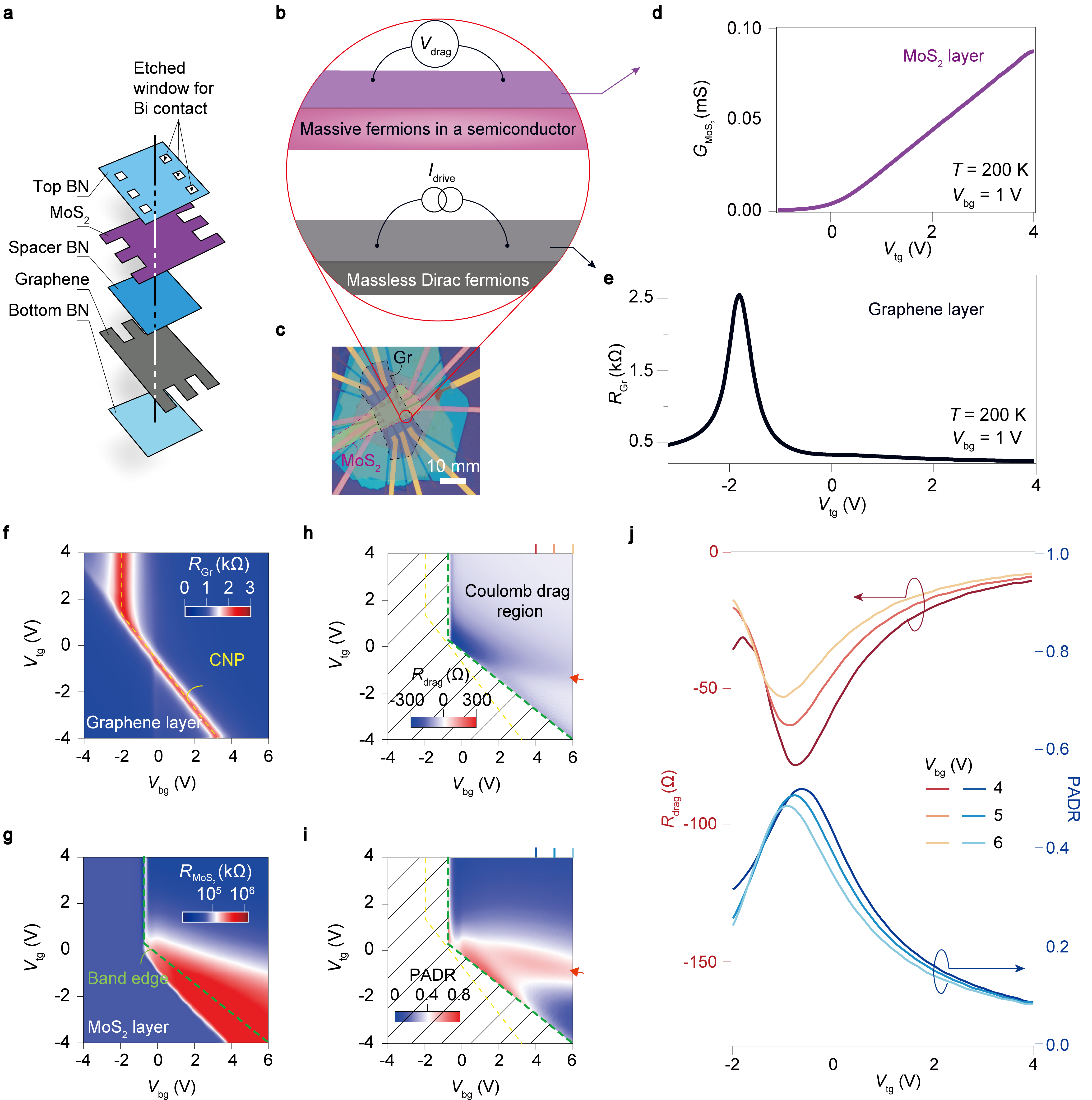}
	\caption{\textbf{Coulomb drag responses in a semiconductor-semimetal hybrid.} (a) Schematic illustration of the heterostructure device in a fashion of vertical assembly. (b) Cartoon drawings of the massive-massless Coulomb drag realized in a MoS$_{2}$–graphene heterostructure separated by an ultrathin 3 nm h-BN dielectric. (c) Optical image of a typical MoS$_{2}$-MLG drag device (Sample S21, bilayer MoS$_{2}$ is used as the semiconducting channel). (d)-(e) Line profile of field effect curves recorded in each constituent layer of graphene and MoS$_{2}$, respectively, with the bottom gate $V_\mathrm{bg}$ = 1 V, and $T$ = 200 K. (f) $R_\mathrm{xx}$ mapping in the $V_\mathrm{tg}$-$V_\mathrm{bg}$ space of the graphene channel in Sample S21. (g) $R_\mathrm{xx}$ mapping in the $V_\mathrm{tg}$-$V_\mathrm{bg}$ space of the MoS$_{2}$ channel. Data obtained at $T$ = 200 K and $B$ = 0 T. (h) Drag responses in the same device. (i) Passive-to-active drag ratio (PADR) for the drag signal tested in the MoS$_{2}$ layer.  Notice that a portion of the map in (h) and (i) are masked (ill defined signal since the lock-in amplifier is out of phase, as seen in Supplementary Figure 6), for visual clarity. (j) The line profiles of $R_\mathrm{drag}$ and PADR at several $V_\mathrm{bg}$ = 4, 5, and 6 V, respectively.}
\end{figure*}

\begin{figure*}[h!]
	\centering
	\includegraphics[width=0.84\linewidth]{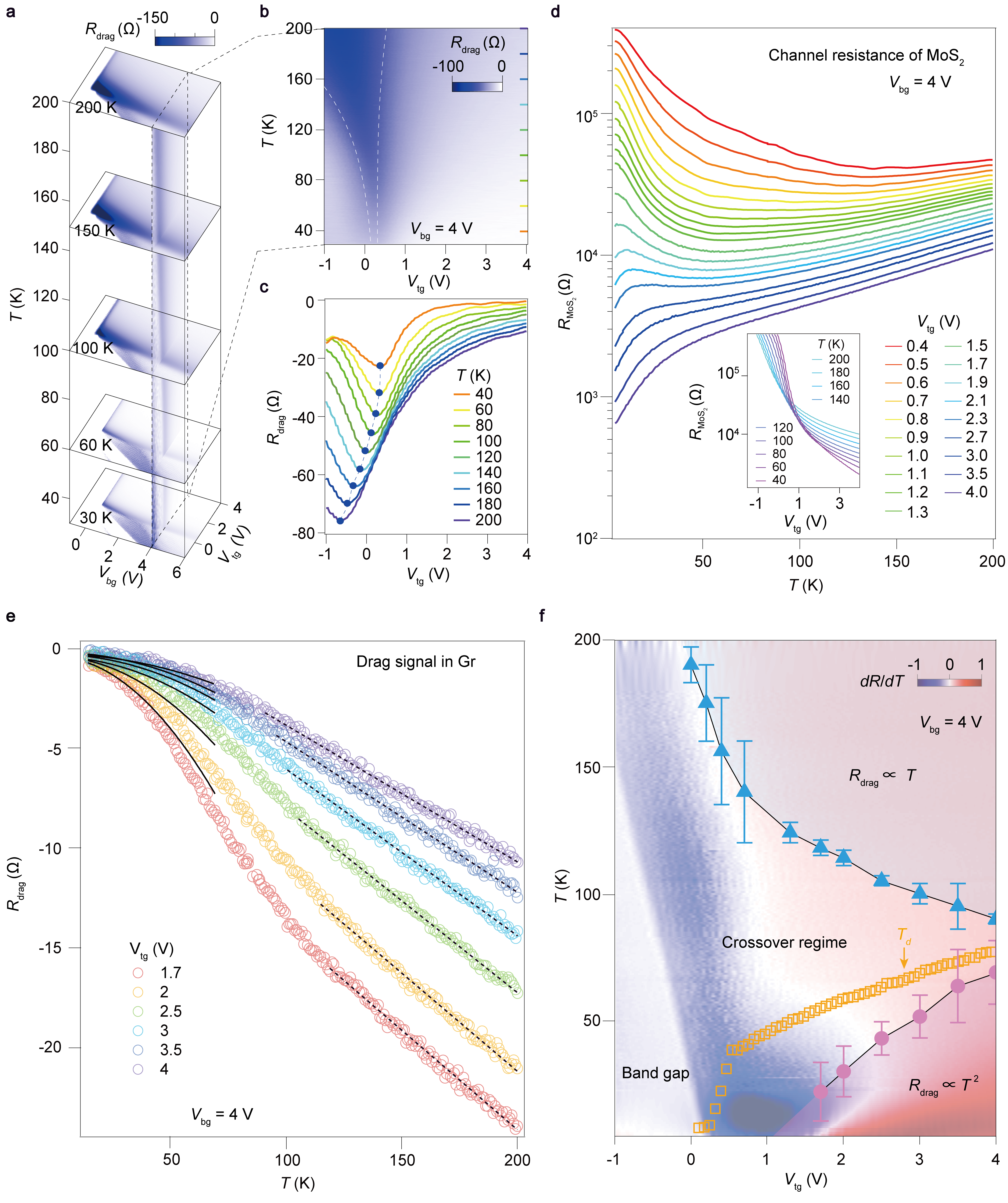}
	\caption{\textbf{Metal-insulator transition in MoS$_{2}$ and the crossover from $T$- to $T^{2}$-dependence of Coulomb drag.} (a) Evolution of $R_{\text{drag}}$ with temperature. Each color map plane depicts  $R_{\text{drag}}$ as a function of $V_{\text{bg}}$ and $V_{\text{tg}}$ at a fixed temperature $T$. The vertical 2D map is $R_{\text{drag}}$ as a function of temperature $T$ and $V_{\text{tg}}$ at $V_{\text{bg}}$ = 4 V, which has been displayed in (b) for further clarification. The white dashed lines in (b) serve as a guide for the eye, indicating the obvious drag signal at different temperatures. (c) The corresponding line cuts of the color map in (b), with the largest drag signal magnitude represented by blue filled circles. As the temperature decreases, the maximum absolute value of $R_{\text{drag}}$ shifts toward lower $V_{\text{tg}}$, indicating a decrease in carrier density. (d) Temperature dependence of the MoS$_{2}$ channel resistance at $V_{\text{bg}} = 4\ V$ for different $V_{\text{tg}}$. The inset shows $R_{\text{MoS}_{2}}$ as a function of $V_{\text{tg}}$ for different temperatures, the data reveal an approximate crossing point at $V_{\text{tg}}$ = 0.5 V. (e) Temperature dependence of $R_{\text{drag}}$ (colored open symbols) at $V_{\text{bg}}$ = 4 V for different $V_{\text{tg}}$. The black solid lines represent fits to the low-temperature data with a quadratic temperature dependence, while the black dashed lines correspond to fits to the high-temperature data, assuming a linear temperature dependence. (f) Temperature–top gate voltage ($T$–$V_{\text{tg}}$) phase diagram of the drag response extracted from our drag measurements in comparison with $dR/dT$ of MoS$_{2}$ at $V_{\text{bg}}$ = 4 V. The blue filled triangles show the boundary below which $R_{\text{drag}}\ (T)$  deviates from the $T$-linear behavior. The purple filled circles show the boundary above which $R_{\text{drag}}\ (T)$  deviates from the $T^{2}$ behavior. The middle region between blue and purple symbols are the $T$-$T^{2}$ dependence crossover regime. The open square symbols are the critical temperature $T_{d}$, which is defined as $T_{d}$ = $E_\mathrm{F}/k_\mathrm{B}k_\mathrm{F}d$ with $k_\mathrm{B}$,$k_\mathrm{F}$ and $d$ being the Boltzmann constant, Fermi vector, and the interlayer distance, respectively. The error bars are defined by the uncertainty in temperature when the difference in $R_{\text{drag}}$ values between the fitting and experimental data are smaller than 0.2 $\Omega$.}
\end{figure*}

\noindent\textbf{Fabrications and characterizations of MLG-MoS$_{2}$ drag devices.} \\ Monolayered graphene, few-layered MoS$_{2}$ and h-BN flakes were mechanically exfoliated from bulk crystals. The MoS$_{2}$ layer is always placed as the upper layer in the drag devices in this study. As illustrated in Fig. 1a, the van der Waals heterostructure is stacked using the dry transfer method \cite{Lei_Science_2013}, and then encapsulated by top and bottom h-BN flakes, with the top h-BN etched into micron-metre sized 2D windows. A windowed contact method is thus deployed to achieve Ohmic contacts to the MoS$_{2}$ channel throughout the temperature range from 5 to 300 K\cite{zhao2024fractional}. The device were equipped with dual metallic gates and electrodes of Ti/Au via standard lithography and electron-beam evaporation (fabrication details are available in Methods). More detailed fabrication processes can be seen in \textcolor{gray}{Supplementary Figures 1-3}. We found that different bottom gate geometry will affect the Coulomb drag measurements (\textcolor{gray}{Supplementary Figure 4}), and the main text will focus on the geometric configuration as illustrated in \textcolor{gray}{Supplementary Figure 1}. 

Figure 1b describes the essential nanostructure in this study -- a semiconductor-semimetal drag hybrid, realized in a MoS$_{2}$–graphene double layer separated by an ultrathin 3 nm h-BN dielectric. Here, considering the low energy physics at the Fermi level within the solid state gate doping range, charge carriers in MoS$_{2}$ and graphene are massive Schrödinger and massless Dirac fermions, respectively. Figure 1c shows the optical micrograph of a typical drag device (Sample S21, in which a bilayer MoS$_{2}$ is utilized), with the corresponding fabrication flow shown in \textcolor{gray}{Supplementary Figure 3}. Within the device, carriers in each layer can be tuned independently. For instance, at $T$ = 200 K, typical field effect curves in the graphene channel (Fig. 1d) and in the MoS$_{2}$ channel (Fig. 1e) can be well obtained, respectively.

\begin{figure*}[h!]
	\centering
	\includegraphics[width=0.85\linewidth]{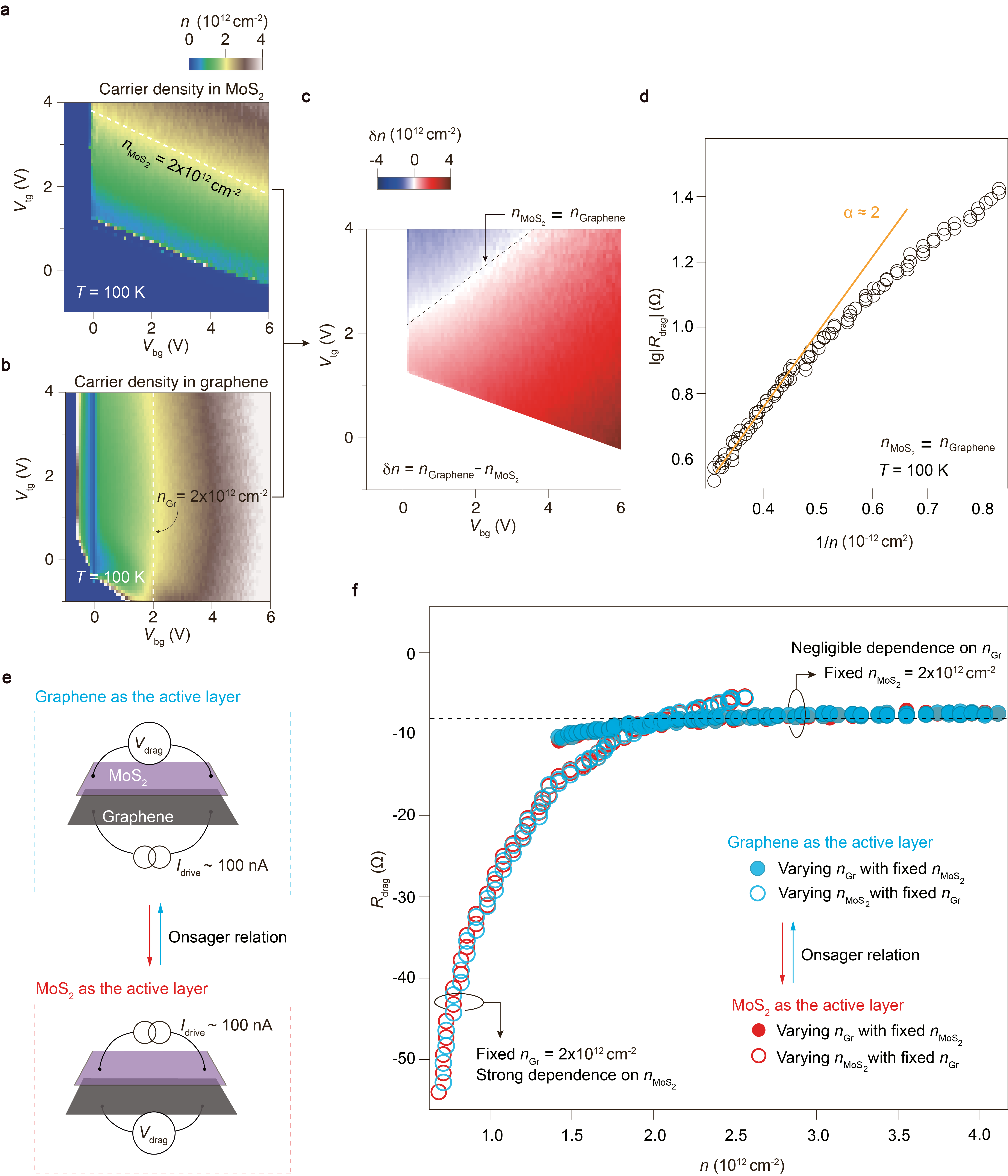}
	\caption{\textbf{Decoupling the density-dependence of massive and massless fermions in $R_{\textrm{drag}}$.} (a)-(b) The carrier density $n$ of each channel in the drag system $R_\mathrm{MoS_{2}}$ and $R_\mathrm{graphene}$ as a function of $V_\mathrm{tg}$ and $V_\mathrm{bg}$. Data are obtained by using the formula $n=B/eR_\mathrm{H}$, where $e$ is the elemental charge and $B/R_\mathrm{H}$ is obtained by extracting the slope of Hall resistance at $B$ = 1 T and 0 T. (c) The differential carrier density $\delta n$ plotted by subtracting the colour map (b) with (a). Notice that black dashed line indicates the scenario of matched-density between the graphene and MoS$_{2}$ layer. (d) $R_{\textrm{drag}}$ plotted alongside the black dashed line in (c), which shows 1/$n^{2}$ dependence in the matched-density drag. (e)  The illustration of the Onsager relation with the active layer alternated in the drag system, while drive current is kept constant at 100 nA. (f) By selectively sweeping the gate voltages in the $V_\mathrm{tg}$-$V_\mathrm{bg}$ space, one can control the variation of solely either $n_\mathrm{graphene}$ (or $n_\mathrm{MoS_{2}}$), with the $n$-dependence decoupled in the drag system. Indeed, $R_{\textrm{drag}}$ is found to have negligible dependence on $n_\mathrm{graphene}$ of massless Dirac fermions, but one order of magnitude stronger dependence on $n_\mathrm{MoS_{2}}$. This dependence is held valid for the Onsager reciprocity relation.}
	\label{fig:fig3}
\end{figure*}

\begin{figure*}[h!]
	\centering
	\includegraphics[width=0.9\linewidth]{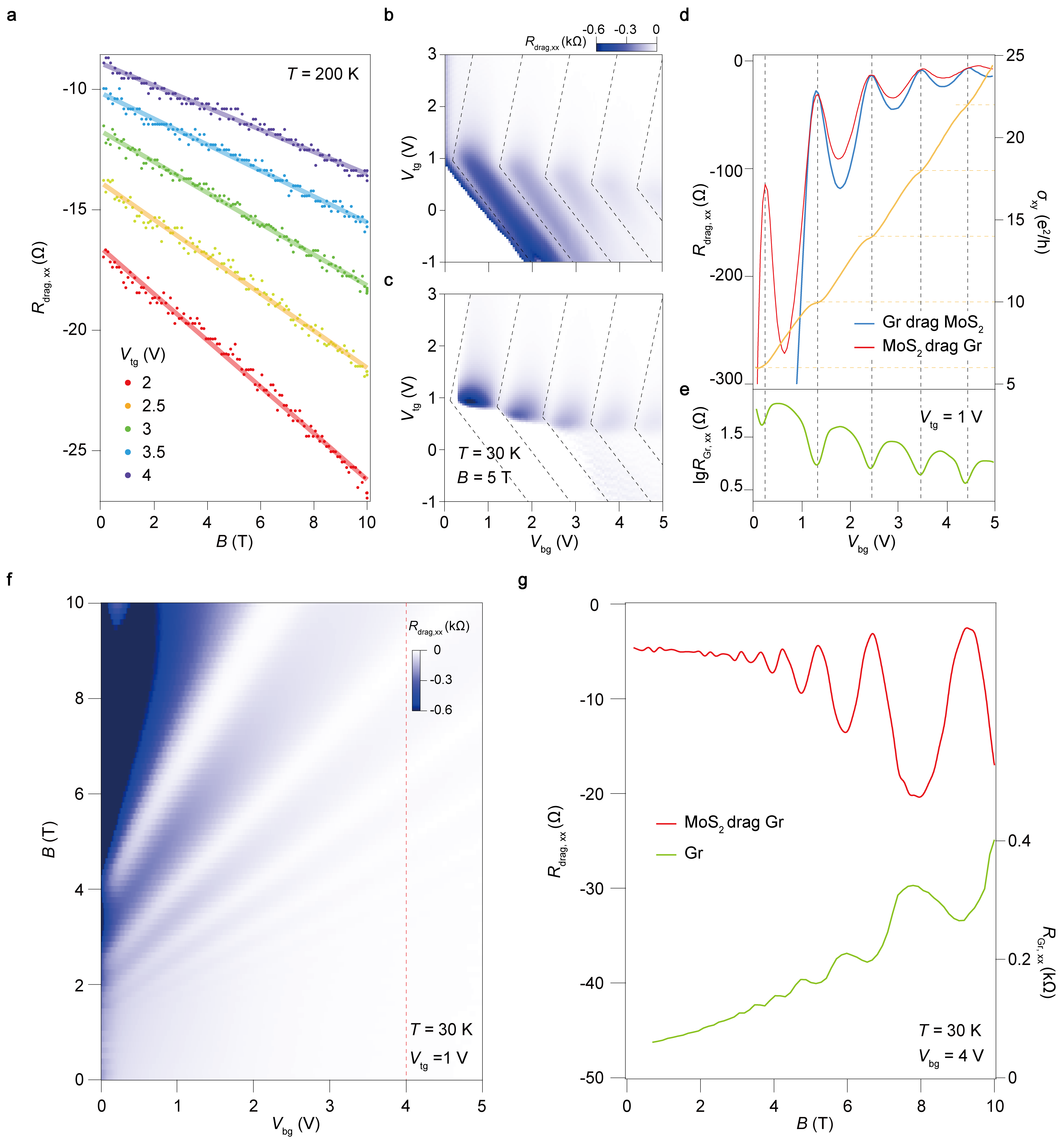}
	\caption{\textbf{Magneto-drag responses in semiconductor-semimetal double layers.} (a) $R_{\textrm{drag}}$ as a function of magnetic field at $T$= 200 K. Linear dependence are seen for several $V_{\textrm{tg}}$. (b)-(c) illustrate the magneto-drag responses in the system, with (b) graphene and (c) MoS$_{2}$ as the active layer, respectively. At lower temperature ($T$= 30 K), Landau levels develop in the graphene layer, and the fulfillment of Onsager relation in the drag system is restricted to the metallic area of MoS$_{2}$. (d) Line profiles of $R_{\textrm{drag}}$ with graphene (blue) and MoS$_{2}$ (red) as the active layer, respectively. (e) Longitudinal channel resistance of graphene along the same gate range as in (d). (f) Landau fan-shaped drag response with graphene as the active layer, in the $B-V_{\textrm{bg}}$ space, while  $V_{\textrm{tg}}$ is fixed at 1 V. (g) Line profile (red solid line) of $R_{\textrm{drag}}$ along the red dashed line indicated in (f), and the corresponding longitudinal channel resistance (solid green line) of the graphene layer.}
	\label{fig:fig4}
\end{figure*}

Figure 1f-g illustrate the mapping of longitudinal channel resistance  $R_\mathrm{graphene}$ and $R_\mathrm{MoS_{2}}$ (in a log scale for visual clarity) in the same $V_\mathrm{bg}$-$V_\mathrm{tg}$ space at $T$ = 200 K in the drag device Sample S21. In general, as shown in Fig. 1f, $R_\mathrm{graphene}$ is in agreement with the previous observation in a standard dual-gated monolayer graphene devices \cite{Pablo_PRL}. However, the charge neutral resistive peak of graphene is partially screened by MoS$_{2}$ due to the existence of relatively high carrier density in the latter layer, yielding a weak $V_\mathrm{tg}$ dependence of $R_\mathrm{graphene}$ at $V_\mathrm{tg}$ larger than $\sim$ 1 V. Meanwhile, the band edge of MoS$_{2}$ in Fig. 1g is squeezed and held almost constant at positive $V_\mathrm{tg}$, which is likely due to the contact part of the MoS$_{2}$ is not gated by the same gate as its major channel. Notice that the band edge of semiconducting MoS$_{2}$ in Fig. 1g is highlighted by green dashed line, which is quantitatively extracted from the phase signal in the lock-in measurement, shown in \textcolor{gray}{Supplementary Figure 5}. 

As a consequence, limited by the screening effect and the contact barriers, drag response in the current device is confined to the electron-electron regime, making hole drag not accessible. For the drag measurements, we passed a drive current ($I_\mathrm{drive}$) through the active layer and measured the resulting voltage drop ($V_\mathrm{drag}$) across the passive layer under open-circuit conditions. To eliminate spurious drag signals in the passive layer caused by drive-bias-induced AC gating effects\cite{hill1996frictional}, we have adopted a balance-bridge setup (comparison between  lock-in measurement and the bridge methods can be seen \textcolor{gray}{Extended Data Figure 1})\cite{liu2017quantum}. The linear relationship between $V_\mathrm{drag}$ and $I_\mathrm{drive}$ confirms the validity of our drag measurements, with the drag resistance $R_\mathrm{drag}$ determined by the slope of the curve (\textcolor{gray}{Extended Data Figure 2}). Figure 1h shows the drag resistance $R_{\textrm{drag}}$ using graphene as the driving layer ($i.e.$, the active layer). Here, the parasitic signals, determined to be coincides with the band edge of MoS$_{2}$ by the phase measurements in \textcolor{gray}{Supplementary Figures 5-6}, are blanked for visual clarity. The results of drag responses are reproducible in different samples, as shown in \textcolor{gray}{Supplementary Figure 7}. The PADR, defined as $I_{\textrm{drag}}/I_{\textrm{drive}}=R_{\textrm{drag}}/R_{\textrm{passive layer}}$, is usually a direct measure of the interlayer interaction in drag systems. For example, when it comes to a perfect drag in the scenario of exciton condensation, PADR may reach the unity\cite{nguyen2023perfect,li2017excitonic,zhang2025excitons}. In our system, PADR (Fig. 1i) has a maximum value of $\sim$ 0.6 when MoS$_{2}$ serves as the active layer, much higher compared to most of the conventional drag systems. Moreover, we notice that maximum $R_{\textrm{drag}}$ seems to take place at the onset of the semiconducting MoS$_{2}$ channel conductance derivative with respect to gate voltage ($dG/{dV_{\mathrm{g}}}$), as indicated by the red arrows in Fig. 1h-i (aslo discussed in \textcolor{gray}{Supplementary Figure 8}). Line profiles of $R_{\textrm{drag}}$ and PADR as a function of $V_\mathrm{tg}$ at typical $V_\mathrm{bg}$ are shown in Fig. 1j. The maximum value of PADR increases as the gate voltage decreases form 6 to 4 V.

\vspace{5mm}
\noindent\textbf{Temperature dependences of the Coulomb drag in the semiconductor-semimetal hybrid.} In the following, we investigate the observed drag response in the graphene-MoS$_{2}$ hybrid at different temperatures. Figure 2a overlays $R_{\textrm{drag}}$ with MoS$_{2}$ being the active layer in the $V_\mathrm{tg}$-$V_\mathrm{bg}$ space at $T$ = 200, 150, 100, 60 and 30 K, respectively. It is seen that the maximum $R_{\textrm{drag}}$ line (alongside the onset of $dG/{dV_{\mathrm{g}}}$ for the MoS$_{2}$ channel, as seen in \textcolor{gray}{Extended Data Figure 3}) is shifted slightly toward higher $V_\mathrm{tg}$ upon lowering the temperature. Indeed, this feature is better captured in the $R_{\textrm{drag}}$ mapping as a function of $T$ and $V_\mathrm{tg}$, as shown in the vertical mapping in Fig. 2a (with MoS$_{2}$ being the active layer) and Fig. 2b (with graphene being the active layer). The line cuts along the colored ticks (which correspond to different temperatures) in Fig. 2b are shown in Fig. 2c. It is clear that a maximum negative $R_{\textrm{drag}}$ is identified in each curve, highlighted by blue solid dots. Discussions on the agreement between the maximum $R_{\textrm{drag}}$ and $dG/{dV_{\mathrm{g}}}$ of the MoS$_{2}$ channel can be seen in \textcolor{gray}{Extended Data Figure 3}.

Interestingly, when lowering the temperature, phonon scattering in the MoS$_{2}$ is known to be largely suppressed and the carrier transport in the system is supposed to be driven from phonon-limited low mobility regime into the intrinsic high mobility regime, with the system exhibiting a transition from insulating behavior to a metallic one, known as metal-insulator transition (MIT) when varying from low to high carrier density in the low temperature limit\cite{radisavljevic2013mobility,moon2018soft,zhao2024fractional}. Indeed, in Sample S21, clear MIT in the MoS$_{2}$ channel can be seen in Fig. 2d, with its inset displaying the channel resistance of MoS$_{2}$ at different temperatures. The linear $I$-$V$ curves of the MoS$_{2}$ channel at different temperatures are plotted in \textcolor{gray}{Supplementary Figure 9}. This highly tunable electron transport properties in one of the layers of the drag system may give rise to unique and unconventional drag signals, distinguishing it from previously reported drag systems\cite{huang2023quantum,wang2024coulomb,pillarisetty2005coulomb,lee2016giant}. Fig. 2e shows the temperature dependence of drag resistance at different $V_\mathrm{tg}$ with the $V_\mathrm{bg}$ fixed at 4 V. $R_{\textrm{drag}}$ increases monotonically as the temperature increases when MoS$_{2}$ becomes metallic at large $V_\mathrm{tg}$. A $T^{2}$ dependence is clearly observed at the base temperatures, which is in good agreement with the theory of frictional drag for Fermi liquid\cite{flensberg1995linear,narozhny2012coulomb}. However, in the high-temperature regime,  deviation from the $T^{2}$ dependence becomes pronounced, eventually evolving into a linear temperature dependence. It is found that the crossover regime from $T^{2}$ to $T$ dependence broadens as MoS$_{2}$ becomes more insulating with decreasing $V_\mathrm{tg}$. 

In order to further analyze the correlation between the drag and pristine transport property of MoS$_{2}$, we compare the color maps of the temperature-dependent resistance deviation ($dR/dT$) of MoS$_{2}$ with the drag resistance in the $T$-$V_\mathrm{tg}$ phase diagram (Fig. 2f). The corresponding fitting points for $R_{\textrm{drag}}$ in Fig. 2e are featured in the phase diagram. From this comparison, we identify four distinct temperature-dependent drag regions. The white region in the color map of $dR/dT$ marks the boundary between the metallic and insulating phases of MoS$_{2}$. On the metallic side of MoS$_{2}$, with the line profiles of the drag resistance already presented in Fig. 2e, clear $T^{2}$, $T$ and $T^{2}$-$T$ crossover regions are observed. Coulomb drag resistance is known to be extremely sensitive to temperature, interlayer spacing, carrier density (or density mismatch between the layers) and magnetic field\cite{li2016negative,zhu2023signature}. And drag transport regimes can be defined by the Fermi energy $E_F$, Fermi momentum $k_F$, interlayer separation $d$. In the Boltzmann-Langevin theory of Coulomb drag, at low temperatures ($T \ll T_{d}$$=E_F/k_Fd$) and in the clean limit (weak disorder or low scattering rate), drag is dominated by the particle-hole continuum and $R_\text{drag}$ is proportional to $T^{2}$\cite{chen2015boltzmann}. Thus, we have plotted the estimated characteristic temperature $T_{d}$ in the phase diagram and found that the curve of $T_{d}$ indeed separates the quadratic $T^{2}$ and linear $T$ dependent drag regimes. The temperature region of a Fermi liquid below $T_{d}$, in which the drag resistance follows the $T^{2}$ law, is strongly suppressed as the MIT boundary is approached. At higher temperatures, $T > T_{d}$, phase-space constraints due to small-angle scattering lead to a linear temperature dependence\cite{chen2015boltzmann,jauho1993coulomb}. While on the insulating side, $R_\text{drag}$ deviates from both $T^{2}$ and linear temperature dependence, eventually drops to zero as the carrier density in MoS$_{2}$ decreases and Fermi level moves into the bandgap of MoS$_{2}$.

\begin{table*}[t!]
	\centering
	\caption{A summary of the characteristics for different drag systems.}
	\resizebox{\textwidth}{!}{%
		\begin{tabular}{|c|c|c|c|c|c|c|}
			\hline
			Drag category & Drag system & \( T \) dependence & \( n \) dependence & \( B \) dependence & Maximum drag resistance & Ref. \\
			\hline 
			Massless-massless fermions 
			& ML Gr-ML Gr   & \( T^2 \) (high density)  & N/A   & anomalous & 50 \( \Omega \) & \cite{gorbachev2012strong} \\
			& ML Gr-ML Gr & \( T^2 \) (0 T)   & N/A  &  \( B^2 \) & 400 \( \Omega \) (70 K, 1 T at CNP) & \cite{liu2017frictional} \\
			\hline
			Massless-massive fermions  
			& ML Gr-CNT  & \( T \) (when \( T > T_\mathrm{F} \))  & \( 1/(V_\mathrm{g}-V_\mathrm{0})^{1\sim 2} \)  & N/A & 6 \( \Omega \) (260 K) & \cite{anderson2021coulomb} \\
			& ML Gr-InAs NW    & \( T^2 \)    & \( 1/n^4 \)  & \( B^2 \) & 0.5 \( \Omega \) (1.5 K) & \cite{mitra2020anomalous}  \\
			& ML Gr-GaAs 2DEG  & \( T^2\ln{T} \) & N/A  & N/A & 2 \( \Omega \) (0.24 K) & \cite{gamucci2014anomalous} \\
			& ML Gr-BL    & \( T^2 \) (high density)   & \( 1/n^2 \) (low density), \( 1/n^3 \) (high density)   & N/A  & 5 \( \Omega \) (high density), 50 \( \Omega \) (CNP) & \cite{zhu2020frictional} \\
			& ML Gr-BL \( \mathrm{MoS_{2}} \) & \( T^2 \sim T \)    & \( 1/n^2 \)  & \( B \)  & 0.3 k\( \Omega \) (200 K, 0 T)& \textbf{This work}  \\
			\hline
			Massive-massive fermions 
			& BL Gr-BL Gr  & \( R_{\mathrm{drag}} \) decreases as \( T \) increases   & N/A  & N/A  & 800 \( \Omega \) (1.5 K at CNP) & \cite{lee2016giant}  \\
			& BL Gr-BL Gr  & \( T^2 \) (nonlocal), \( T^4 \) (local) & \( 1/n^3 \) (nonlocal, low density) & N/A & 60 \( \Omega \) (CNP) & \cite{li2016negative}  \\
			& ML \( \mathrm{MoSe_{2}} \)-ML \( \mathrm{WSe_{2}} \)  & \( T^2 \) (<10 K) & \( 1/(n^\mathrm{p}-n^\mathrm{m})^3 \) & N/A & 1 M\( \Omega \) (1.5 K) & \cite{nguyen2023perfect}  \\
			& FL \( \mathrm{MoS_{2}} \)-FL \( \mathrm{MoS_{2}} \)  & \( T^2\ln{T} \) & N/A & N/A & 2.5 M\( \Omega \) (1.5 K) & \cite{huang2023quantum}  \\
			& BL Gr-GaAs 2DEG  & N/A  & \( 1/n^3 \) (high density)  & N/A  &  2 \( \Omega \) (70 K) & \cite{simonet2017anomalous} \\
			\hline
			Others 
			& ML Gr-LAO/STO & \( T_c \sim \) 0.2 K & N/A & N/A & 0.5 \( \Omega \) (0.2 K) & \cite{tao2023josephson} \\
			& InAs-GaSb topological wires & \( R_{\mathrm{drag}} \) decreases as \( T \) increases & N/A & N/A & 0.8 k\( \Omega \) (0.3 K) & \cite{du2021coulomb} \\
			& ML Gr-Gr/BN moiré superlattice & \( T^2 \) (high density) & \( 1/n^{1.3 \sim 1.7} \) (high density) & N/A & 10 \( \Omega \) (CNP) & \cite{wang2024coulomb} \\
			\hline
		\end{tabular}%
	}
\end{table*}

\bigskip
\noindent\textbf{Drag at the matched density.} It is noticed that the carrier density dependent characteristic of $R_\mathrm{drag}$ varies significantly in different kinds of drag system.  The relationship between drag resistance and carrier density at the matched density ($n_{\mathrm{MoS_2}}$ = $n_{\mathrm{Graphene}}$) in massive-massless fermion system has been explored theoretically, which is in contrast with that in massive-massive and massless-massless fermion systems\cite{hwang2011coulomb,scharf2012coulomb}. For high density regime ($k_Fd\ \gg \ 1$), all three systems exhibit a similar carrier density dependence, specifically following an $1/n^{3}$ behavior. For the low density regime ($k_Fd\ \ll \ 1$), the carrier density dependence exhibits distinct characteristics for different systems, highlighting their unique properties. Specifically, in the massless-massive case, $R_\mathrm{drag}$ scales as $1/n^{2}$, whereas for massive-massive and massless-massless systems, the dependencies are predicted to follow $1/n^{3}$ and $1/n$, respectively\cite{narozhny2012coulomb,scharf2012coulomb}. In our case, we first estimate $n_{\mathrm{MoS_2}}$ and $n_{\mathrm{Graphene}}$ independently from the longitudinal and transverse resistance ($R_\mathrm{xx}$ and $R_\mathrm{xy}$, respectively) of the MoS$_{2}$ and graphene layers based on measurements of the classical Hall effect at 100 K, as shown in Fig. 3a and b. The $R_\mathrm{xy}$ of both MoS$_{2}$ and graphene varies linearly with the magnetic field, as shown in \textcolor{gray}{Supplementary Figure 10}.  The equal density line ($n_{\mathrm{MoS_2}}$ = $n_{\mathrm{Graphene}}$) can be easily identified by subtracting the two carrier density color maps of MoS$_{2}$ and graphene, as illustrated by the black dash line in Fig. 3c. Subsequently, we plot the drag resistance along the density matched line ($n_{\mathrm{MoS_2}}$ = $n_{\mathrm{Graphene}}$) in logarithmic scale and converges to the expected $1/n^{\alpha}$ dependence, with $\alpha \approx2$. For our massive-massless fermion system, the range of the equal density line is about $1.2 \sim 3.0 \times10^{12}\, \mathrm {cm}^{-2}$. Thus, we estimate that the maximum value of $k_Fd$ satisfies $k_Fd < 1$ using the expression $k_F = \sqrt{\pi n}$. This result demonstrates the equal density line is at the low density regime with the $1/n^{2}$ dependent $R_\mathrm{drag}$, which agrees well with the theoretical predicts\cite{scharf2012coulomb}. 
\\
\indent 
In addition to the matched density condition, we also explore the density-dependent drag resistance in the non-equal density case ($n_{\mathrm{MoS_2}} \neq n_{\mathrm{Graphene}}$) to examine the varying capabilities of controlling the drag signal by the two different layers. In the following analysis, we selectively sweep the top and bottom gate voltages, in order to adjust the carrier density of MoS$_{2}$ (or graphene) independently while keeping the density of graphene (or MoS$_{2}$) constant, and consider the Onsager reciprocal relations (illustrated schematically in the measurement configuration of Fig. 3e) of the resulted drag responses. We use the same density of MoS$_{2}$ or graphene, fixed at $2\times10^{12} \ \mathrm {cm}^{-2}$ (indicated by the white dashed lines in Fig. 3a-b for $n_{\mathrm{MoS_2}}$ and $n_{\mathrm{Graphene}}$, respectively) for comparison. Noticeably, the drag resistance exhibits a pronounced dependence on $n_{\mathrm{MoS_2}}$, while showing a weak dependence on $n_{\mathrm{Graphene}}$. The validity of the Onsager reciprocal relations is confirmed for the decoupled density-dependent drag results, as evidenced by the agreement between the blue and red symbols, where the active layer is graphene and MoS$_{2}$, respectively. In both cases, variations in only $n_{\mathrm{MoS_2}}$ are indicated by solid circles, while variations in only $n_{\mathrm{graphene}}$ are indicated by open circles. Note that the results in Fig. 3f are not consistent with the drag response reported for the MLG-BLG massive-massless fermion system, where the drag resistance follows the functional dependence $R_{\text{drag}} = f(n_t + n_b)$ or the conventional $R_{\text{drag}} = f(n_t \times n_b)$, where $n_{\mathrm{t}}$ and $n_{\mathrm{b}}$ are the carrier densities of top and bottom layers\cite{simonet2017anomalous,zhu2020frictional,narozhny2016coulomb}. Our results demonstrate that, in 2D semicondutor-graphene drag system, the control ability of $n_{\text{Graphene}}$ over the drag signal may be one order of magnitude smaller than that of $n_{\text{semiconductor}}$, and $R_{\text{drag}} \approx f(n_{\text{semiconductor}})$ applies under certain conditions. It may be originated from the much stronger electron correlations in the semiconductor MoS$_{2}$ compared to the semimetallic graphene.

\bigskip
\noindent\textbf{Magneto-drag in the MoS$_{2}$-graphene hybrid.} Finally, we show the magnetodrag (the longitudinal component) of massive-massless fermion system in the presence of a finite magnetic field $B$. We first investigate the $B$-dependent $R_{\text{drag}}$ at 200 K, when graphene and MoS$_{2}$ are not in the quantum hall regime as shown in Fig. 4a. Remarkably, $R_{\text{drag, xx}}$ does not exhibit the conventional frictional drag $B^{2}$ dependence but instead shows a linear $B$ behavior at different top gate voltages from $V_{\text{tg}}$ = 2 to 4 V, accompanied by the linear $T$-dependence observed at high temperatures (Fig. 2e-f). When the temperature is further lowered to 30 K, the graphene layer enters the quantum Hall state with integer Landau level (LL) filling fractions at 5 T (see \textcolor{gray}{Supplementary Figure 11}), whereas MoS$_{2}$ doesn't. Consequently, magnetodrag $R_{\text{drag, xx}}$ in both Fig. 4b and 4c show well-developed stripped features (highlighed by the broken lines), which are regions of nearly zero drag responses. The reciprocal magnetodrag $R_{\text{drag, xx}}$ along with the longitudinal resistance and transverse conductance of graphene at 30 K and 5 T are clearly shown in Fig. 4d and 4e, respectively. The Onsager reciprocity is still valid and the observed minimum absolute values of oscillations in magnetodrag signal are consistent with the gapped states between LLs in graphene. Therefore, the vanishing drag signals arise from inefficient drag due to the insulating and incompressible nature of the graphene bulk, which leads to a vanishing density of states for interlayer Coulomb scattering\cite{tse2019magneto,liu2017frictional}. This phenomenon becomes more pronounced when the magnetic field becomes larger (see \textcolor{gray}{Supplementary Figures 12-13}). It is noteworthy that Onsager reciprocal relation is only valid when MoS$_{2}$ becomes metallic. Fig. 4f shows the fan diagram of magnetodrag $R_{\text{drag, xx}}$ as a function of $V_\text{bg}$ at 30 K, with $\mathrm{MoS_{2}}$ as the active layer. The fan diagram of magnetodrag is similar to that obtained in the pristine graphene channel and held valid for the Onsager reciprocity relation, as shown in \textcolor{gray}{Supplementary Figure 14}. The absolute magnitude of the drag signal increases with the magnetic field and the smallest measured magnitudes of oscillations in the magnetodrag signal align with the presence of gaps between Landau levels in graphene, as displayed in Fig. 4g. Shown in Table 1, characteristics of temperature, magnetic field, and carrier density dependence in a collection of experimentally tested Coulomb drag systems\cite{lee2016giant, gorbachev2012strong, liu2017frictional, anderson2021coulomb, mitra2020anomalous, gamucci2014anomalous, zhu2020frictional, nguyen2023perfect, huang2023quantum, li2016negative,simonet2017anomalous, tao2023josephson, du2021coulomb, wang2024coulomb} are summarized. Compared to those reported, the massless Dirac -massive Schrödinger fermions graphene-MoS$_{2}$ drag system in this work demonstrates an unconventional crossover from $T^2$ to $T$-dependence (Fig. 2e-f), as well as a linear magneto-drag response (Fig. 4a), providing a distinct paradigm for future theoretical considerations.

\bigskip
\noindent To conclude, by introducing  semiconducting TMD channel with Ohmic contacts, we have devised a drag system consisting of graphene–MoS$_{2}$ heterostructure separated by an ultrathin h-BN dielectric. It demonstrates the emergence of a large Coulomb drag response, along with a transition from linear to quadratic temperature dependence of the drag resistance, accompanied by the metal-insulator transition in MoS$_{2}$. The experimental platform enables precise control over intralayer interaction-driven drag by independently tuning carrier densities in each layer, offering new insights into the interplay between massive Schrödinger and massless Dirac carriers. The dominance of a large Wigner–Seitz radius ($r_{s}$>10) in MoS$_{2}$ indicates that electron correlations play a crucial role in shaping the drag response, with graphene acting as a passive layer. Furthermore, a linear magneto-drag response was observed in the  graphene–MoS$_{2}$ heterostructure drag devices, distinguishing it from previously known systems. Our findings enrich the drag family and suggest that a semiconductor-semimetal double-layer 2D electronic system may be intriguing for the design of unique interaction physics in Coulomb drag charge transports.

\section*{Methods}
\vspace{3mm}
\noindent\textbf{Sample fabrication.} vdW few-layers of the h-BN/$\text{MoS}_{2}$/h-BN/graphene/h-BN sandwich were obtained by mechanically exfoliating high quality bulk crystals. The vertical assembly of vdW layered compounds were fabricated using the dry-transfer method in a nitrogen-filled glove box. The heterostructures were then transferred onto the pre-fabricated Au or graphite gates. Hall bars of the devices were achieved by reactive ion etching. During the fabrication processes, electron beam lithography was done using a Zeiss Sigma 300 SEM with a Raith Elphy Quantum graphic writer. One-dimensional edge contacts of monolayer graphene were achieved by using the electron beam evaporation with Ti/Au thicknesses of $\sim$ 5/50 nm and the window contacts of bilayer $\text{MoS}_{2}$ were fabricated with a thermal evaporator, with typical Bi/Au thicknesses of $\sim$ 25/30 nm. After atomic layer deposition of about 20 nm Al$_{2}$O$_{3}$, big top gate was deposited to form the complete dual gated h-BN encapsulated drag devices as shown in Fig. 1a and c.

\vspace{3mm}
\noindent\textbf{Drag measurements.} In lock-in measurements, current is typically driven by applying an AC bias voltage $V_{\text{drive}}$ to one side of the channel while the other side is grounded. However, in Coulomb drag measurements, applying this bias to the drive layer may induce spurious drag signals in the drag layer due to the AC gating effect caused by the drive bias. Here we applied about 0.2 $\sim$ 0.3 V AC bias voltage at 17.777 Hz to drive the active layer through a 1:1 voltage transformer. The transformer was connected to a 10 k$\Omega$ potentiometer, which can help to distribute the AC voltage across both ends of the driving layer. This configuration minimizes the AC interlayer potential difference in the channel, thereby reducing the AC coupling between the active and passive layers. We used two 1 M$\Omega$ resistors connected with the driving layer and measured the voltage drop across one of the resistors to obtain the driving current. The drag voltages were recorded using low-frequency SR830 lock-in amplifiers. Four-probe measurements were used throughout the transport measurements in an Oxford Teslatron cryostat. Gate voltages on the as-prepared devices were controlled by a Keithley 2400 source meter.

\section*{\label{sec:level1}Data Availability}

The data that support the findings of this study are available upon reasonable request to the corresponding authors.

\section*{\label{sec:level2}Code Availability}

The code that support the findings of this study are available upon reasonable request to the corresponding authors.

\section*{\label{sec:level3}Acknowledgements}
This work is supported by theNational Key R$\&$D Program of China (Grant Nos. 2024YFA1410400, and 2022YFA1203903) and the National Natural Science Foundation of China (NSFC) (Grant Nos. 12450003, and 92265203). Z.H. acknowledges the support of the Fund for Shanxi “1331 Project” Key Subjects Construction, and supports from the Innovation Program for Quantum Science and Technology (Grant No. 2021ZD0302003). K.W. and T.T. acknowledge support from the JSPS KAKENHI (Grant Numbers 20H00354 and 23H02052) and World Premier International Research Center Initiative (WPI), MEXT, Japan.

\section*{Author Contributions}
S.Z., Z.H., and W.R. conceived the experiment and supervised the overall project. Y.L. and S.Z. performed the device fabrications and electrical measurements; K.Y. contributed to electrical measurements; K.W. and T.T. provided high quality h-BN bulk crystals; S.Z., Y.L. and Z.H. analysed the experimental data. The manuscript was written by Z.H., S.Z. and Y.L. with discussions and inputs from all authors.

\section*{Competing Interests}
The authors declare no competing interests.

\clearpage
\newpage

\renewcommand{\thefigure}{1}
\renewcommand{\figurename}{Extended Data Fig.}
\begin{figure*}[t!]
	\centering
	\includegraphics[width=0.9\linewidth]{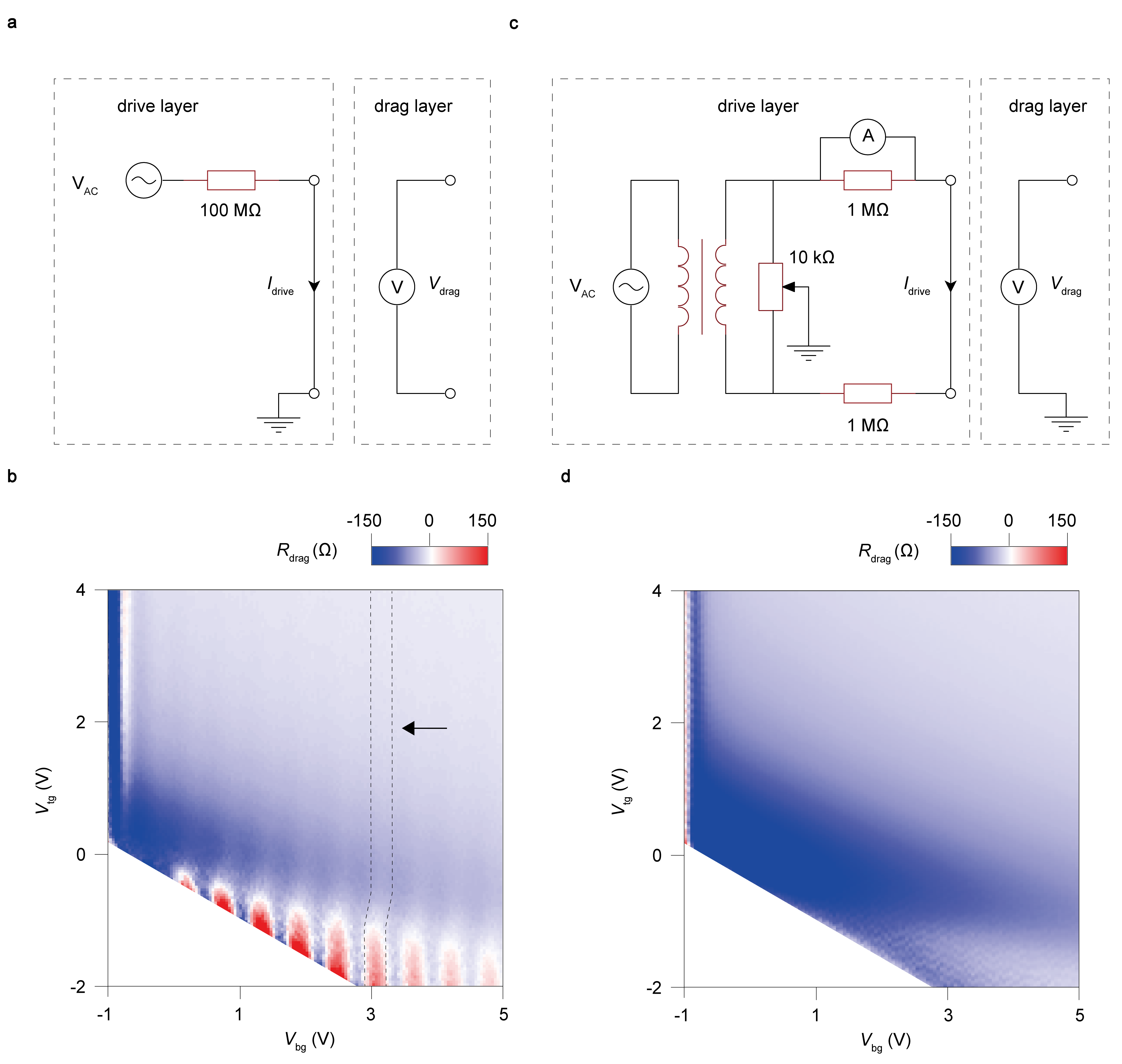}
	\caption{\textbf{Two kinds of setup for measuring Coulomb drag signals.} (a) The circuit for a simple lock-in measurement. A current is driven through one layer, while the resulting voltage is measured across the other layer, which remains open-circuited. As a result, the potential at the midpoint of the drive layer reaches $\sim$ \({V_{\text{drive}}}/2\) with respect to ground. Given that the drag layer is grounded, an AC interlayer bias of $\sim$ \({V_{\text{drive}}}/2\) is generated accordingly and produce spurious drag signal, as shown by the dashed line in (b). (c) The circuit for the Coulomb drag measurement. The AC voltage is fed into the bridge circuit through a 1:1 ground-isolating transformer to minimize ground loop. By carefully tuning the variable resistor in the bridge, the AC potential at the center of the drive layer is adjusted to approximately zero, preventing interlayer capacitance coupling and ensuring accurate measurement of the drag signal. The 2D map of the corresponding drag resistance as a function of $V_\mathrm{bg}$ and $V_\mathrm{tg}$ is shown in (d).}
    \label{fig:Extended Data Fig.1} 
\end{figure*}

\clearpage
\newpage

\renewcommand{\thefigure}{2}
\renewcommand{\figurename}{Extended Data Fig.}
\begin{figure*}[t!]
	\centering
	\includegraphics[width=0.9\linewidth]{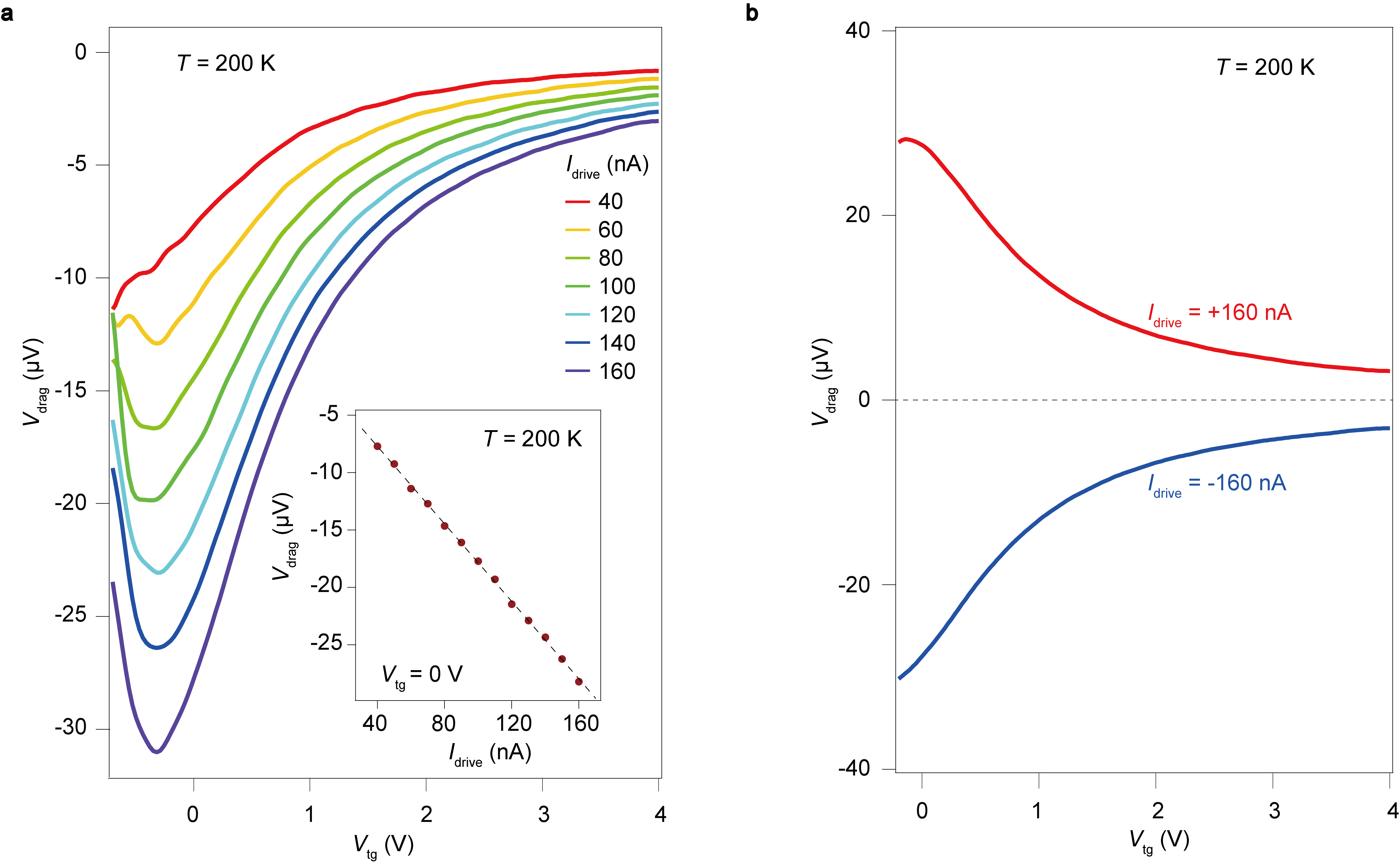}
	\caption{\textbf{The relation between $V_{\text{drag}}$ and $I_{\text{drive}}$.} (a) The measured $V_{\text{drag}}$ as a function of $V_\mathrm{tg}$ at $V_\mathrm{bg}$ = 1 V and $T$ = 200 K for different drive currents. The inset shows the extracted values of $V_{\text{drag}}$ at $V_\mathrm{tg}$ = 0 V as a function of $I_{\text{drive}}$. The black dashed line represents the linear fit, demonstrating the linear response of the drag signal to $I_{\text{drive}}$. (b) The nearly symmetric drag response in different directions of drive current at $T$ = 200 K.}
    \label{fig:Extended Data Fig.2} 
\end{figure*}

\clearpage
\newpage

\renewcommand{\thefigure}{3}
\renewcommand{\figurename}{Extended Data Fig.}
\begin{figure*}[ht!]
	\centering
	\includegraphics[width=0.9\linewidth]{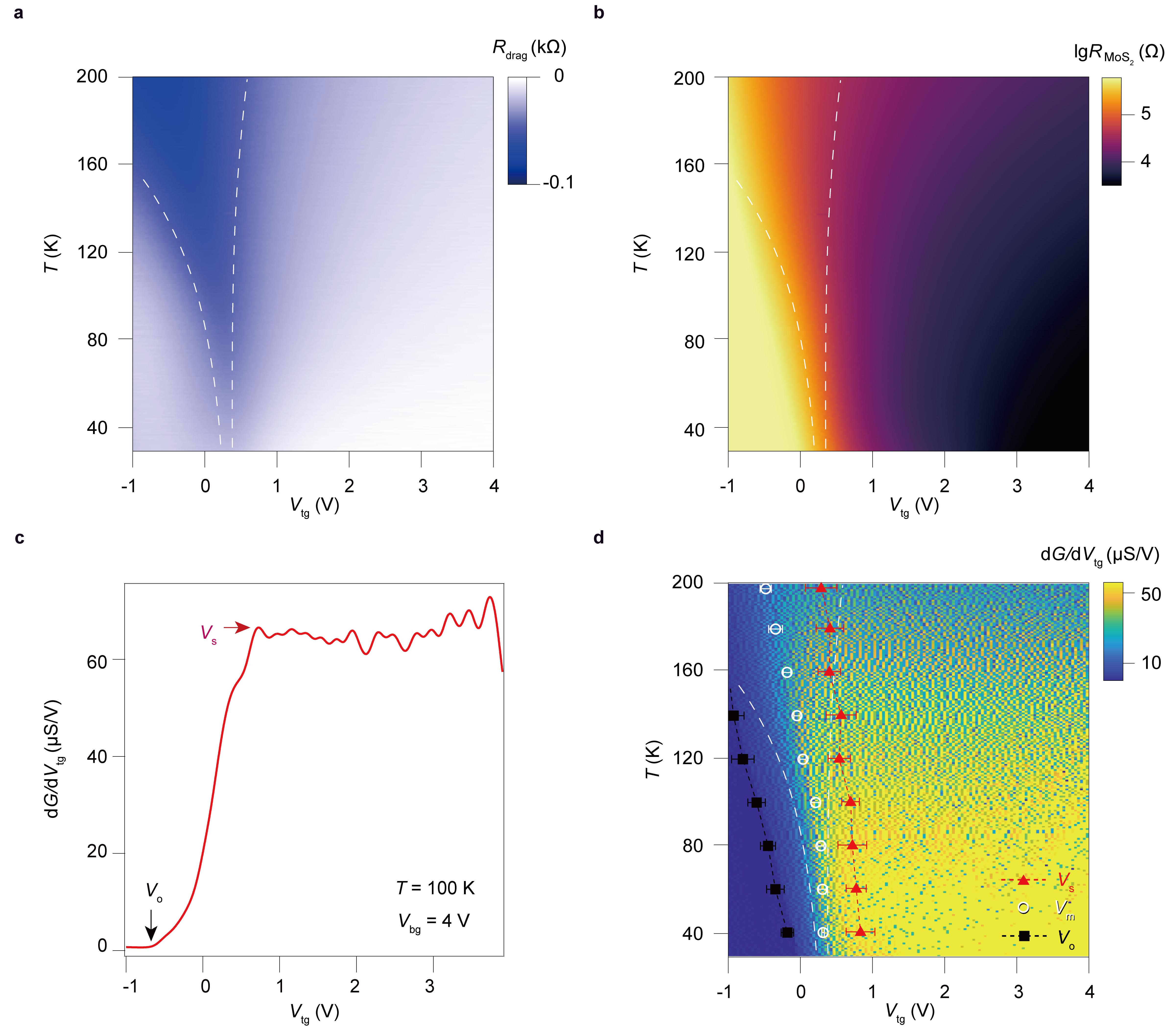}
	\caption{\textbf{Comparison of the drag resistance with the channel resistance and $dG/{dV_{\mathrm{g}}}$ of $\mathrm{MoS_2}$}. 2D map of (a) drag resistance and (b) channel resistance of $\mathrm{MoS_2}$ as a function of temperature and $V_\mathrm{tg}$. (c) The line profile of $dG/{dV_{\mathrm{g}}}$ for $\mathrm{MoS_2}$ as a function of $V_\mathrm{tg}$ at $T$ = 100 K and $V_\mathrm{bg}$ = 4 V. The onset point $V_\mathrm{o}$ and the saturate point $V_\mathrm{s}$ of $dG/{dV_{\mathrm{g}}}$ for $\mathrm{MoS_2}$ are indicated by black and red arrows, respectively.
    (d) 2D map of $dG/{dV_{\mathrm{g}}}$ for $\mathrm{MoS_2}$ as a function of $T$ and $V_\mathrm{tg}$. The black square and red triangle filled symbols represent $V_\mathrm{o}$ and $V_\mathrm{e}$ at different temperatures, respectively. The error bars are determined by taking 1\micro S/V of the onset values of $dG/{dV_{\mathrm{g}}}$ and 90\% of the saturation magnitude at different temperatures, respectively. The white open symbols represent the largest magnitude position of drag signal $V_\mathrm{m}$. The error bars are determined by 1\% of the largest magnitude of $V_\mathrm{m}$. From this figure, we can clearly see that the largest magnitude of drag signal happens in the middle of $V_\mathrm{o}$ and $V_\mathrm{e}$, corresponding to the region between the onset and saturation of $dG/{dV_{\mathrm{g}}}$.}
    \label{fig:Extended Data Fig.3} 
\end{figure*}

\clearpage
\newpage

\renewcommand{\thefigure}{4}
\renewcommand{\figurename}{Extended Data Fig.}
\begin{figure*}[t!]
	\centering
	\includegraphics[width=0.9\linewidth]{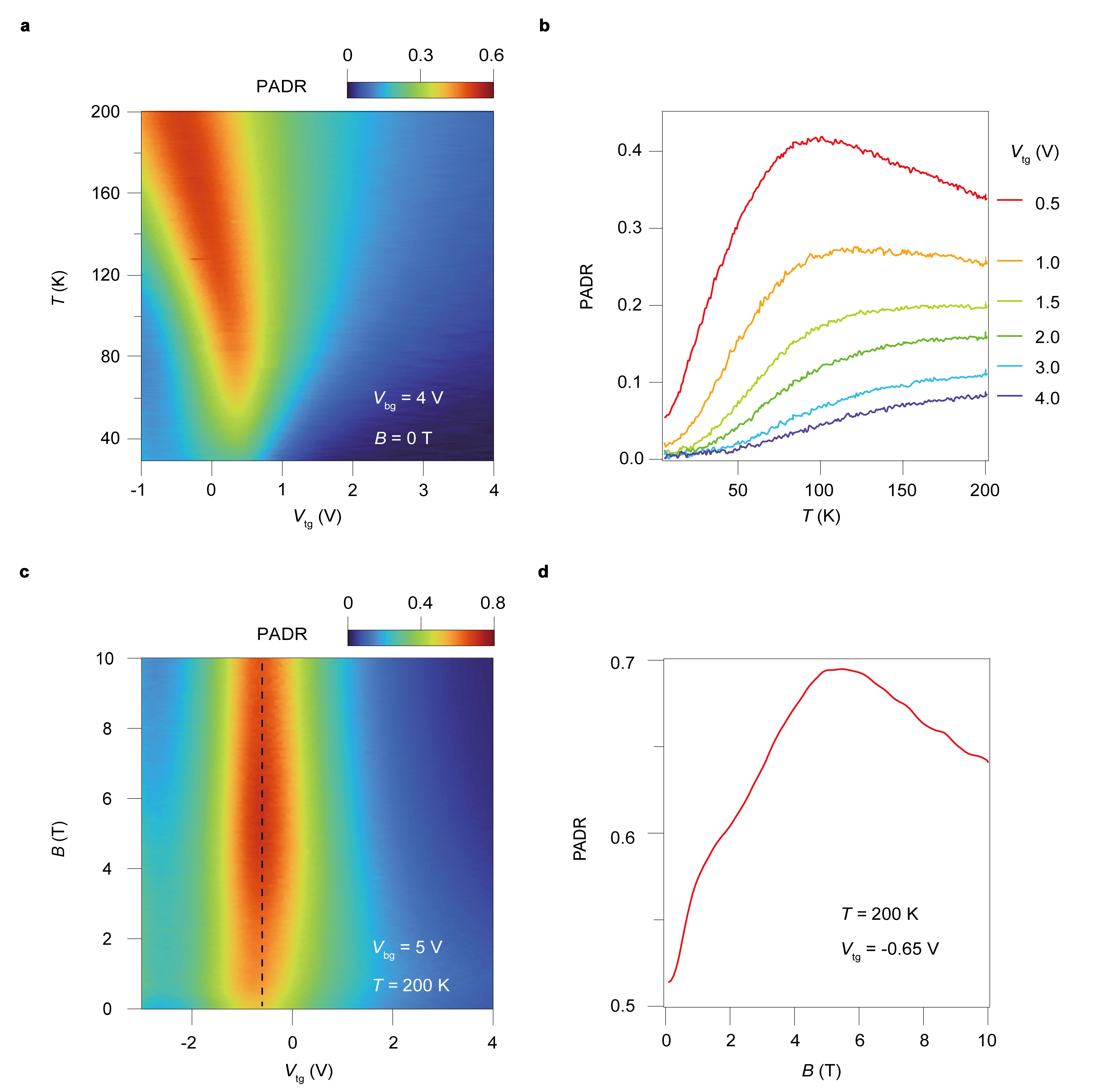}
	\caption{\textbf{Tuning PADR by adjusting magnetic field and temperature.} (a) The 2D map of PADR as a function of $V_\mathrm{tg}$ and $T$ at $V_\mathrm{bg}$ = 4 V. (b) The temperature dependence of PADR at different $V_\mathrm{tg}$. (c) The 2D map of PADR as a function of $V_\mathrm{tg}$ and $B$ at $V_\mathrm{bg}$ = 5 V and $T$ = 200 K. (d) The magnetic-field dependence of PADR at $V_\mathrm{tg}$ = -0.65 V is shown, as indicated by the black dashed line in panel (c). The magnetic field can effectively modulate PADR by approximately 40\%, reaching a maximum value of around 0.7.}
    \label{fig:Extended Data Fig.4} 
\end{figure*}

\end{document}